\documentstyle[twoside,fleqn,espcrc2]{article}

\newcommand{\AmS}{{\protect\the\textfont2
  A\kern-.1667em\lower.5ex\hbox{M}\kern-.125emS}}

\newcommand{\Tr}{{\rm Tr}}
\newcommand{\Dmrns}{{\cal D}_{\mu\rho,\nu\sigma}}

\title{ 
\hfill\begin{minipage}{0pt}\scriptsize\vspace*{-1.5cm} \begin{tabbing}
\hspace*{\fill} IFUP-TH 34-98
\end{tabbing} 
\end{minipage}\\[-8pt]
Gauge--invariant nonlocal quark condensates in QCD}

\author{M. D'Elia\thanks{Speaker at the conference.}\address{Department of 
            Natural Sciences, University of Cyprus, PO Box
            537 Nicosia, CY-1678, Cyprus}, 
        A. Di Giacomo\address{Dipartimento
                  di Fisica, Universit\`a di Pisa, Piazza Torricelli 2,
                  56126 Pisa, Italy},
        E. Meggiolaro\address{Institut f\"ur Theoretische Physik, 
         Universit\"at Heidelberg, Philosophenweg 16, D-69120 
         Heidelberg, Germany}}
\begin{document}

\begin{abstract}
We study, by numerical simulations on a lattice, the
behaviour of the gauge--invariant nonlocal quark condensates 
in the QCD vacuum both in the {\it quenched} approximation and
with four flavours of dynamical 
staggered fermions. The correlation length of the condensate
is determined to be roughly twice as big as in the case of
the gluon field strength correlators.

\end{abstract}

\maketitle

\section{INTRODUCTION}

The QCD sum rules approach  to  
non--perturbative phenomena in strong
interaction physics consists in evaluating
the power corrections of the product of two hadronic currents via the 
Operator Product Expansion (OPE), putting the relevant, unknown information
about the non-perturbative vacuum structure in the values of local
condensates ($\langle \bar{q}(0) q(0) \rangle$, 
$\langle G(0) G(0) \rangle$, etc.).
However, it has been recognized in~\cite{Gromes82,Mik-Rad86,Rad91} that in many applications
the $x$-distribution of the vacuum fields cannot be safely neglected:
this leads to the study of gauge--invariant field correlators, or ``nonlocal condensates''.

The gauge--invariant two--point correlators of the gauge field strengths
in the QCD vacuum, defined as:
\begin{equation}
\Dmrns(x) = \langle 0| 
\Tr \left\{ G_{\mu\rho}(0) S G_{\nu\sigma}(x) S^\dagger \right\}
|0\rangle ~,
\end{equation}
where  $G_{\mu\rho} = gT^aG^a_{\mu\rho}$ is the field--strength tensor and
$S = S(x,0)$ is the Schwinger phase operator,
have been extensively studied on the lattice in the past 
\cite{Campostrini84,DiGiacomo92,npb97,plb97}. 
The basic result is that the correlator
$\Dmrns(x)$, in the Euclidean theory, can be written as the sum of a 
perturbative--like term, behaving as $1/|x|^4$, and a non--perturbative
part, which falls down exponentially
\begin{equation}
\Dmrns^{(n.p.)}(x) \sim \exp ( -|x|/\lambda_A ),
\end{equation}
with a correlation length $\lambda_A \simeq 0.13$ fm for the quenched $SU(2)$ 
theory \cite{Campostrini84}, $\lambda_A \simeq 0.22$ fm for the 
quenched $SU(3)$ theory \cite{DiGiacomo92,npb97} and $\lambda_A \simeq 0.34$
fm for full QCD (approaching the chiral limit) \cite{plb97}.

Along the same line, in this paper we present a lattice determination
of the quark--antiquark nonlocal condensates, defined as:
\begin{equation}
C_i (x) = -\displaystyle\sum_{f=1}^4
\langle \Tr [\bar{q}^f_a (0) (\Gamma^i)_{ab} S(0,x) q^f_b (x)] \rangle ~,
\end{equation}
where $S(0,x)$ is the Schwinger line
needed to make $C_i (x)$ gauge--invariant and  $\Gamma^i$ are the
sixteen independent  matrices of the Clifford's algebra
acting on the Dirac indices $a,b$.
The trace in (3) is taken with respect of the colour indices.

The computations have been performed 
both in the {\it quenched} approximation and in full QCD using 
four degenerate flavours of {\it staggered}
fermions (whence the sum over the flavour index $f$ in (3))
and the $SU(3)$ Wilson action for the pure--gauge sector.
We have adopted the same basic strategies and techniques already 
developed for the study of the gluonic correlators.

In full QCD the nonlocal 
condensates have been measured on a 
$16^3 \times 24$ lattice at $\beta = 5.35$ and 
two different values of the quark mass: 
$a \cdot m_q = 0.01$ and $a \cdot m_q = 0.02$. 

For the {\it quenched} case the measurement has been performed on a $16^4$
lattice at $\beta = 6.00$, using a valence quark mass $a \cdot m_q = 0.01$, 
and at $\beta = 5.91$ with a  quark mass $a \cdot m_q = 0.02$. 
The $\beta$ values have been
chosen in order to have the same physical scale as in full QCD
at the corresponding quark masses, thus allowing a direct comparison 
between the {\it quenched} and the full theory.
In this way it is possible to estimate the effects of the 
inclusion of dynamical fermions on the quantities under study.
In the {\it quenched} case two other measurements have been done at 
$\beta = 6.00$, using valence quark masses $a \cdot m_q = 0.05$ and $0.10$.

Further details about the computation and the results will soon be 
published~\cite{nostro98}.

\begin{table*}[ht]
\setlength{\tabcolsep}{5.0pc}
\newlength{\digitwidth} \settowidth{\digitwidth}{\rm 0}
\catcode`?=\active \def?{\kern\digitwidth}
\caption{Results for the correlation length 
$\lambda_0$. Reported errors
refer only to our determination and do not include the uncertainty on 
the physical scale (``$f$'' stands for {\it ``full--QCD''}, while
``$q$'' stands for {\it ``quenched''}).} 
\label{tab:tab}
\begin{tabular*}{\textwidth}{@{}l@{\extracolsep{\fill}}rrrr}
\hline
&                   \multicolumn{1}{r}{$\beta$, theory} 
                 & \multicolumn{1}{r}{$a \cdot m_q$} 
                 & \multicolumn{1}{r}{$\lambda_0$ (fm)}         \\
\hline
&5.35, f & 0.01 & $0.63^{+0.21}_{-0.13}$ \\
&5.35, f & 0.02 & $0.46^{+0.09}_{-0.06}$ \\
&6.00, q & 0.01 & $0.64^{+0.22}_{-0.13}$ \\
&5.91, q & 0.02 & $0.46^{+0.06}_{-0.05}$ \\
&6.00, q & 0.05 & $ 0.30(2) $      \\
&6.00, q & 0.10 & $ 0.187(3)$      \\
\hline
\end{tabular*}
\end{table*}

\begin{figure}[t]
\vspace{3.41cm}
\includegraphics{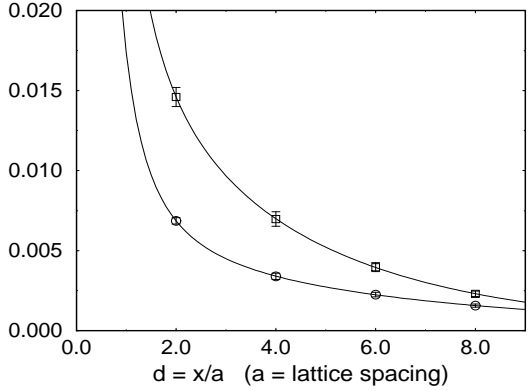}
\null\vskip 0.15cm
\caption{The function $a^3 C_0(x)$ 
versus the distance $d$, for the full--QCD case at
$\beta = 5.35$ and quark masses $a \cdot m_q = 0.01$ (circles) and
$a \cdot m_q = 0.02$ (squares). 
The curves correspond to our best fits [Eq. (6)].}
\end{figure}

\begin{figure}[t]
\vspace{3.41cm}
\includegraphics{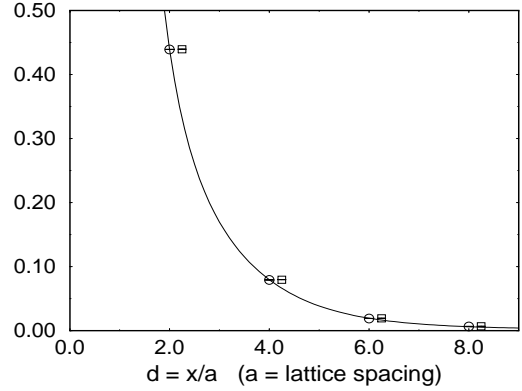}
\null\vskip 0.15cm
\caption{The function $a^3 C_v(x)$
versus the distance $d$, for the full--QCD case at
$\beta = 5.35$ and quark masses $a \cdot m_q = 0.01$ (circles) and
$a \cdot m_q = 0.02$ (squares). 
The curve corresponds to our best fit [Eq. (6)] to the data
at $a \cdot m_q = 0.01$.}
\end{figure}

\section{COMPUTATION AND RESULTS}

Making use of T,P invariance one 
can prove that all the correlators (3) vanish, except 
those with $\Gamma^i = {\bf 1}$ ({\it ``scalar nonlocal condensate''})
and with  $\Gamma^i = \gamma_E^\mu$ and $\mu$ in the direction of $x$ 
({\it ``longitudinal-vector nonlocal condensate''}):
\begin{eqnarray}
C_0 (|x|) = -\displaystyle\sum_{f=1}^4 \langle \Tr [ \bar{q}^f_a (0)
S(0,x) q^f_a (x) ] \rangle ~; \nonumber \\
C_v (|x|) =
-{x_\mu \over |x|}  \displaystyle\sum_{f=1}^4 \langle \Tr [ \bar{q}^f_a (0)
(\gamma^\mu_E)_{ab} S q^f_b (x) ] \rangle ~.
\end{eqnarray}

To discretize our operators on the lattice we have 
properly combined the staggered propagators in order to build up 
the physical quark propagators. The Schwinger 
phase operator, i.e., the straight line of links 
connecting the point $0$ to the point $x$, has then been put in.
In this way we construct two lattice operators
$C^L_0 (d \cdot a)$ and $C^L_v (d \cdot a)$ ($x = d \cdot a$, where $d$ is 
the distance in units of lattice spacings $a$), which are 
proportional, in the
na\"{\i}ve continuum limit, to $C_0 (d \cdot a)$ and $C_v (d \cdot a)$ respectively:
\begin{eqnarray}
C^L_0 (d \cdot a) \mathop\sim_{a\to0} a^3 
C_0 (d \cdot a) + {\cal O}(a^4) ~,\nonumber \\
C^L_v (d \cdot a) \mathop\sim_{a\to0} a^3 
C_v (d \cdot a) + {\cal O}(a^4) ~.
\end{eqnarray}
Higher orders in $a$ in (5) as well as possible multiplicative 
renormalizations are removed by cooling the quantum fluctuations at the scale 
of the lattice spacing, as explained in 
Refs.~\cite{DiGiacomo92,npb97,plb97,Campostrini89,DiGiacomo90}. 
This removal will show up as a plateau in the dependence of the correlators 
on the number of steps of the cooling procedure: our data are the values of 
the correlators at the plateaux.

The nonlocal condensates have been measured
at distances $d = 2,4,6,8$.
In Figs. 1 and 2 we display the results for $a^3 C_0 (d \cdot a)$ and
$a^3 C_v (d \cdot a)$ respectively obtained in full QCD.

We have tried a best fit to 
the data with the following functions:
\begin{eqnarray}
C_0 (x) &=& A_0 \exp (-\mu_0 x) + {B_0 \over x^2} ~; \nonumber \\ 
C_v (x) &=& A_v x^3 \exp (-\mu_v x) + {B_v \over x^3} ~.
\end{eqnarray}

The form of the perturbative--like terms in Eq. (6) (i.e.,
$B_0/x^2$ for the scalar condensate and $B_v/x^3$ for the vector 
condensate) is that obtained in the leading order in perturbation theory, in 
the chiral limit $m_q \to 0$. Further details, as well
as a remark about the reliability of the results obtained 
for the  vector nonlocal condensate,
will be soon published~\cite{nostro98}.

\section{DISCUSSION}

\noindent
The quantity of greatest physical interest which can be extracted from our 
lattice determinations is the correlation length 
$\lambda_0 \equiv 1 / \mu_0$
of the scalar quark--antiquark nonlocal condensate. 
It plays a relevant role in many applications of QCD sum rules, 
especially for studying the pion form factors and the pion wave 
functions~\cite{Mik-Rad86,Rad91}.
In Table I we report the values obtained for $\lambda_0$ in all 
the cases examined. The physical scale in the full QCD case has been set
by measuring $m_\pi$ and $m_\rho$ on our configurations, while for the 
{\it quenched} case  we have used results reported in Ref. \cite{boyd}. 

At $a \cdot m_q = 0.01$ the value is roughly twice as big as the value for 
the correlation length $\lambda_A$ of the gluon field strength at the same quark mass \cite{plb97}.
As in the gluon case \cite{plb97}, the fermionic correlation length 
appears to decrease when increasing the quark mass. However, 
the full--QCD and the {\it quenched} values of $\lambda_0$ are 
nearly the same, when compared at the same quark mass, thus suggesting
that this quantity is not 
much influenced by the introduction of dynamical quark loops.

Finally, using the values of the pion mass $m_\pi$ 
measured on our configurations
in the full--QCD case 
we find that $m_\pi \lambda_0 = 1.6(4)$ for $a \cdot m_q = 0.01$ and
$m_\pi \lambda_0 = 1.4(2)$ for $a \cdot m_q = 0.02$. Therefore, within the
errors, the inverse of the scalar correlation length, $\mu_0 = 1/\lambda_0$,
turns out to be proportional to the pion mass $m_\pi$.

\section*{Acknowledgements}
Partially supported by EC TMR Program
ERBFMRX-CT97-0122, and by MURST, project: ``Fisica Teorica delle
Interazioni Fondamentali''.

\end{document}